\documentstyle[preprint,aps,epsf]{revtex}

\newcommand {\bea}{\begin{eqnarray}}
\newcommand {\eea}{\end{eqnarray}}
\newcommand {\be}{\begin{equation}}
\newcommand {\ee}{\end{equation}}

\begin{document}

\preprint{SUNY-NTG-01-03}

\title{Possible Color Octet Quark-Anti-Quark Condensate 
in the Instanton Model}

\author{Thomas Sch\"afer}

\address{Department of Physics, SUNY Stony Brook,
Stony Brook, NY 11794\\ and\\
Riken-BNL Research Center, Brookhaven National 
Laboratory, Upton, NY 11973}

\maketitle

\begin{abstract}

 Inspired by a recent proposal for a Higgs description of 
QCD we study the possible formation of a color-octet/flavor-octet 
quark-anti-quark condensate in the instanton liquid model. For this 
purpose we calculate two-point correlation functions of color-singlet 
and octet quark-anti-quark operators. We find long range 
order in the standard $\langle\bar{\psi}\psi\rangle$
channel, but not in the color-octet channel. We 
emphasize that similar calculations in lattice QCD
can check whether or not a color-flavor locked 
Higgs phase is realized in QCD at zero temperature
and baryon density.

\end{abstract}

\vspace*{1cm}

 It was recently argued that confinement in QCD can be understood 
in terms of an effective Higgs description where $SU(3)$ gauge invariance 
is spontaneously broken by a color-octet quark-anti-quark condensate
\cite{Wetterich:1999vd,Wetterich:2000pp,Wetterich:2000ky,Berges:2000dh}
(see \cite{Srednicki:1981cu,Iijima:1981tt} for earlier speculations 
in this direction)
\be
\label{oct}
\langle \bar{\psi}(\lambda^A)_{C}(\lambda^A)^T_F\psi\rangle
 = 2\left( \delta^a_i\delta^b_j-\frac{1}{3}\delta^{ab}
  \delta_{ij} \right) \langle \bar\psi^a_i \psi^b_j \rangle .
\ee
Here, $(\lambda^A)_C$ ($A\!=\!1,\ldots,8$) is a color $SU(3)$ matrix 
and $(\lambda^A)_F$ is a flavor $SU(3)$ matrix. Color indices in the 
fundamental representation of $SU(3)$ are denoted by $a,b$ and 
flavor indices by $i,j$.  The condensate (\ref{oct}) breaks 
the local gauge symmetry as well as the $SU(3)_L\times SU(3)_R$ 
chiral symmetry but leaves a diagonal $SU(3)$ unbroken. This symmetry
acts as the physical flavor symmetry. The Higgs field (\ref{oct})
allows for a remarkably simple description of the gross
features of the QCD spectrum. Gluons acquire a mass via the 
Higgs mechanism and carry the flavor quantum numbers and 
charges of the vector meson octet. The quark degrees of 
freedom transform as an octet and a heavier singlet and 
carry the quantum numbers of the baryon octet and singlet. 

 This type of Higgs description was originally proposed as
a complementary picture of QCD at large baryon chemical potential
\cite{Alford:1999mk,Schafer:1999ef}. In that case, the primary Higgs 
field is given by the color-flavor locked diquark condensate
\be
\label{cfl}
\langle \psi_i^a C\gamma_5 \psi_j^b\rangle
 = \phi \left(\delta_i^a\delta_j^b-\delta_i^b\delta_j^a\right).
\ee
This condensate has the same residual color-flavor symmetry 
as (\ref{oct}) but also breaks the $U(1)$ of baryon number. At 
large baryon density this is not a concern and simply corresponds 
to baryon superfluidity. The diquark condensate is dynamically 
generated by attractive interactions in the color anti-triplet 
quark-quark channel. At very large chemical potential this attraction 
is generated by one-gluon exchange \cite{Bailin:1984bm} while at 
intermediate density instanton induced interactions are likely 
to play a role \cite{Rapp:1998zu,Alford:1998zt}. Because the 
color-octet quark-anti-quark condensate (\ref{oct}) is consistent 
with the symmetries of the color-flavor locked diquark phase we 
expect this operator to have a non-zero expectation value in the high 
density phase \cite{Rapp:2000qa,Schafer:2000fe,Pisarski:2000gq}. 
This expectation is borne out by explicit calculations 
\cite{Rapp:2000qa,Schafer:2000fe} but the value of the 
color-octet condensate is strongly suppressed with
respect to the primary Higgs field, $\langle\bar{\psi}(\lambda^A)_C
(\lambda^A)_F^T\psi\rangle \simeq \langle\bar{\psi}\psi\rangle 
\ll \langle\psi C\gamma_5(\lambda^A)_C(\lambda^A)_F\psi\rangle$. 
The suppression is due to an approximate $Z_2$ symmetry 
in the high density phase. 
 
 At zero baryon density the dynamical origin of a possible
color-octet quark-anti-quark is unclear. Both one-gluon exchange 
and instantons are repulsive in this channel. Nevertheless,
given the attractiveness of the scenario outlined in 
\cite{Wetterich:1999vd,Wetterich:2000pp,Wetterich:2000ky,Berges:2000dh}
it seems worthwhile to investigate this question beyond
perturbation theory. In this note, we report on a calculation using 
the instanton liquid model \cite{Schafer:1998wv,Diakonov:1995ea}. 
This model correctly accounts for chiral symmetry breaking
at zero baryon density and the formation of a diquark condensate
at high density. The color-octet quark-anti-quark condensate at 
high density is also induced by instantons. Furthermore, Wetterich 
suggested that an instability in the instanton induced effective 
potential in three flavor QCD might be the dynamical origin of the 
color-octet quark-anti-quark condensate \cite{Wetterich:2000ky}.

 First we have to discuss how to identify the Higgs phase
characterized by (\ref{oct}). The color-octet condensate (\ref{oct}) 
is not a gauge invariant operator and therefore it cannot be used 
as an order parameter. The simplest alternative is the square
of the octet condensate. This is similar to studying the vev
of the square of the $SU(2)$ Higgs field in the standard
model. In the present case, however, this is not very 
useful. The square of the octet condensate receives large
contributions from fluctuations associated with ordinary
chiral symmetry breaking. This is suggested by the 
factorization approximation \cite{Shifman:1979bx}
which gives
\be
\label{fact}
 \langle \left(\bar{\psi}(\lambda^A)_{C}(\lambda^A)^T_F
  \psi\right)^2\rangle \simeq
  \frac{8}{81}\langle\bar\psi\psi\rangle^2 .
\ee
In the instanton model $\langle (\bar{\psi}(\lambda^A)_{C}
(\lambda^A)^T_F\psi)^2\rangle$ is about 2-3 times 
larger than this estimate, but even larger deviations from
factorization are observed in other Lorentz-scalar four-quark
operators. Instead of the square of the octet condensate
we propose to study the color-octet quark-anti-quark 
correlation function. If QCD can be described in terms
of the Higgs picture advocated in \cite{Wetterich:1999vd}
we expect to observe long range order in this correlation 
function. We should note that the color-octet correlation 
function is also not a gauge invariant object. This problem 
can be addressed by calculating the correlator in some fixed
gauge or by inserting two gauge strings. One might worry that 
if the gauge strings are included the correlator will no longer 
approach a constant even if there is long range order. Nevertheless, 
if the Higgs description makes sense there has to be a clear 
difference between the color-octet correlator and a generic gauge 
invariant correlation function without long range order. 

 We have calculated the correlation functions in the 
color-singlet/flavor-singlet, color-singlet/flavor-octet 
and color-octet/flavor-octet channel. The 
correlation functions are defined by
\bea
\label{11}
\Pi_{[1][1]}(x,y) &=&  -\langle 
 {\rm Tr}\left[S^{ab}(x,y)S^{ba}(y,x)\right]\rangle
+ \langle 
 {\rm Tr}\left[S^{aa}(x,x)\right]
 {\rm Tr}\left[S^{bb}(y,y)\right]\rangle, \\
\label{18}
\Pi_{[1][8]}(x,y) &=& - \langle 
 {\rm Tr}\left[S^{ab}(x,y)S^{ba}(y,x)\right]\rangle, \\
\label{88}
\Pi_{[8][8]}(x,y) &=& -4\langle 
 {\rm Tr}\left[S^{aa}(x,y)S^{bb}(y,x)\right]\rangle
 +\frac{4}{3}\langle 
 {\rm Tr}\left[S^{ab}(x,y)S^{ba}(y,x)\right]\rangle,
\eea
where $S^{ab}(x,y)$ is the quark propagator with 
color indices $a,b$ and the traces are taken over 
Dirac indices. We have assumed exact flavor 
symmetry. The flavor-singlet correlation function
$\Pi_{[1][1]}$ is strongly attractive and expected
to approach $\langle\bar\psi\psi\rangle^2$ as $x-y$
tends to infinity. The flavor-octet correlator 
$\Pi_{[1][8]}$ is expected to decay exponentially
with a characteristic mass $m_{[1][8]}\simeq m_{a_0}
\simeq 1$ GeV. Depending on whether the Higgs 
description is valid we expect the color-flavor octet 
correlator $\Pi_{[8][8]}$ to behave similar to 
$\Pi_{[1][1]}$ or to $\Pi_{[1][8]}$.

 Figures 1 and 2 show the three correlation functions
(\ref{11}-\ref{88}) measured in quenched and unquenched 
instanton liquid simulation. In both cases we observe  
long range order in the $\bar{\psi}\psi$ channel,
$\Pi_{[1][1]}(x)\to const$. From the asymptotic value
of the correlation function we find $\langle \bar\psi
\psi\rangle \simeq-(250\,{\rm MeV})^3$ in the 
quenched case and $-(220\,{\rm MeV})^3$ in the 
unquenched case. The flavor-octet $\Pi_{[1][8]}$ 
correlation function decays exponentially. In the 
quenched calculation the correlator becomes unphysical
for $x>0.5$ fm. In the unquenched calculation the 
screening mass is consistent with 1 GeV. 

 We observe that the color-flavor octet correlator 
$\Pi_{[8][8]}$ also decays exponentially. There is no 
sign of long range order. In the unquenched calculation
the screening mass is 700 MeV, while it is even larger
in the quenched calculation. We have also calculated 
the correlator with the gauge links included. We observe 
no qualitative difference. In fact, the screening mass
is slightly increased.


 Finally, we would like to address another potential 
concern regarding our method for studying the scenario
proposed in \cite{Wetterich:1999vd}.
The color-octet condensate (\ref{oct}) breaks no physical
symmetries except for the chiral $SU(3)_L\times SU(3)_R$ 
symmetry. It does, however, break the original flavor symmetry 
of the theory. The unbroken $SU(3)_V$ is a linear combination
of the original $SU(3)_V$ and $SU(3)_C$ symmetries. This might
imply that the color-octet condensate (\ref{oct}) cannot be
observed in a finite volume unless a 
flavor-symmetry breaking source term is added. The proper 
way to extract the color-octet condensate is then to take 
the thermodynamic limit first and set the source term 
to zero afterwards. In order to investigate this possibility
we have calculated the color-octet/flavor-octet correlation
function in the case of unequal quark masses $m_u,m_d,m_s$. 
This means that the flavor symmetry is completely broken. The
color-octet/flavor-octet correlation function is given by
\bea
\Pi_{[8][8]}(x,y) &=& 4\left\{
   \sum_{f,g}
      \langle\left( {\rm Tr}\left[S^{ff}_f(x,x)\right]
       -\frac{1}{3} {\rm Tr}\left[S^{aa}_f(x,x)\right] \right)
       \left( {\rm Tr}\left[S^{gg}_g(y,y)\right]
       -\frac{1}{3} {\rm Tr}\left[S^{aa}_g(y,y)\right] \right)
    \rangle \right. \nonumber \\
 & & \mbox{} - \sum_{f,g}   
     \langle {\rm Tr}\left[S^{ff}_f(x,y)S^{gg}_g(y,x)\right]\rangle
   +\frac{2}{3} \sum_f
     \langle{\rm Tr}\left[S^{fa}_f(x,y)S^{af}_f(y,x)\right]\rangle
   \nonumber \\
\label{88_fsb}
 & & \mbox{} \left.-\frac{1}{9} \sum_f 
  \langle{\rm Tr}\left[S^{ab}_f(x,y)S^{ba}_f(y,x)\right]\rangle
   \right\} 
\eea
Here, $f,g$ are flavor indices, $S^{ab}_f=\{S^{ab}_{u},S^{ab}_{d},
S^{ab}_s\}$, $S^{fa}_f=\{S^{1a}_{u},S^{2a}_{d},S^{3a}_s\}$
and $S^{ff}_f=\{S^{11}_u,S^{22}_d,S^{33}_s\}$. It is 
easy to check that in the case of exact flavor symmetry, $S^{ab}_u
=S^{ab}_d=S^{ab}_s$, expression (\ref{88_fsb}) reduces to our 
earlier result (\ref{88}). We note that the correlation function
(\ref{88_fsb}), unlike the correlator in the flavor-symmetric case
(\ref{88}), contains disconnected contributions. This makes it 
more likely to observe long range order. In Figure 3 we show 
the color-octet/flavor-octet correlation function for different
values of the flavor-symmetry breaking parameter $\delta m$.
This parameter is related to the masses of the up, down and
strange quarks by $m_u=m_0$, $m_d=m_0+\delta m$ and $m_s
=m_0+2\delta m$. Results are shown for $\delta m=0.0,0.1,0.2,
0.3\,{\rm fm}^{-1}$ where $m_0=0.1\,{\rm fm}^{-1}$ is kept
fixed. We observe that the color-octet/flavor-octet correlation 
function increases with $\delta m$, but again there is no evidence 
for long range order. There is also no clear evidence for
a non-analytic dependence on $\delta m$. 

 We should note that a similar analysis can be carried 
out for the color-singlet/flavor-octet correlation
function $\Pi_{[1][8]}$. In this case the appearance
of long range order for $\delta m\neq 0$ would 
correspond to spontaneous flavor symmetry breaking
and $\langle\bar{u}u-\bar{d}d\rangle \neq 0$. This 
possibility is excluded by the Witten-Vafa theorem. 
In agreement with this theorem, no spontaneous
flavor symmetry breaking is observed in the 
instanton model. 

 In summary we observe no evidence for long 
range order in the color-flavor octet channel. 
We suggest that lattice calculations of the
correlation functions (\ref{11}-\ref{88_fsb}) can
provide a definitive answer to the question
whether the Higgs picture suggested in 
\cite{Wetterich:1999vd} is realized in nature.

Acknowledgements: This work was supported in part by US DOE grant 
DE-FG-88ER40388. I would like to thank C. Wetterich and 
E. Shuryak for useful discussions.

\newpage\noindent

\begin{figure}
\begin{center}
\leavevmode
\vspace{1cm}
\epsfxsize=10cm
\epsffile{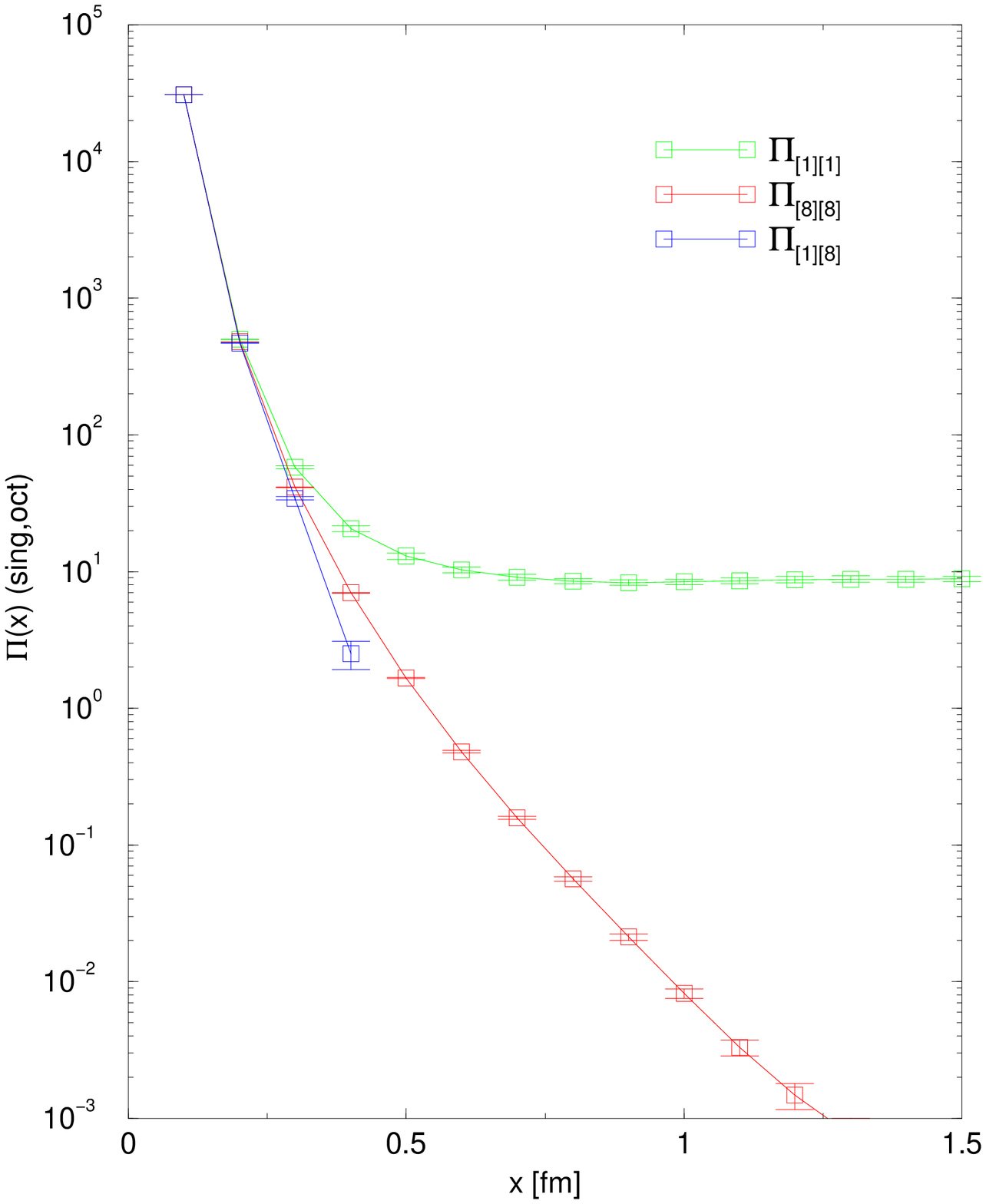}
\end{center}
\caption{Correlation functions in the color-singlet/flavor-singlet
([1][1]), color-octet/flavor-octet ([8][8]) and 
color-singlet/flavor-octet ([1][8]) channel. The correlators were 
calculated in a quenched instanton ensemble. The function 
$\Pi_{[8][8]}$ was multiplied by a factor $3/32$ in order to 
normalize the short distance behavior of all correlation functions 
in the same way.}
\end{figure}

\begin{figure}
\begin{center}
\leavevmode
\vspace{1cm}
\epsfxsize=10cm
\epsffile{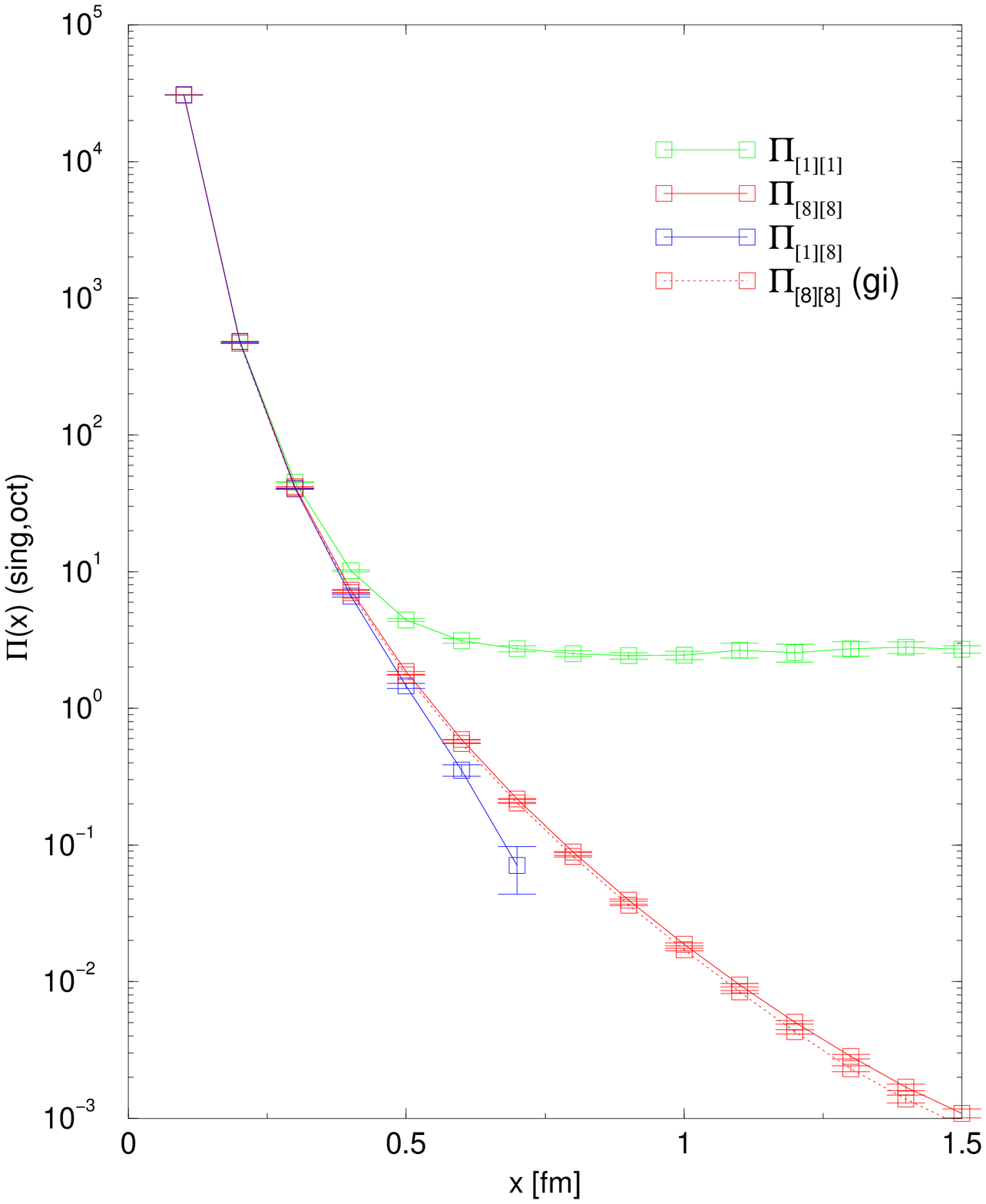}
\end{center}
\caption{Same correlators as in Figure 1 calculated in an 
unquenched instanton ensemble. $\Pi_{[8][8]}({\rm gi})$
denotes the color-octet correlator with the gauge links
included.}
\end{figure}

\begin{figure}
\begin{center}
\leavevmode
\vspace{1cm}
\epsfxsize=10cm
\epsffile{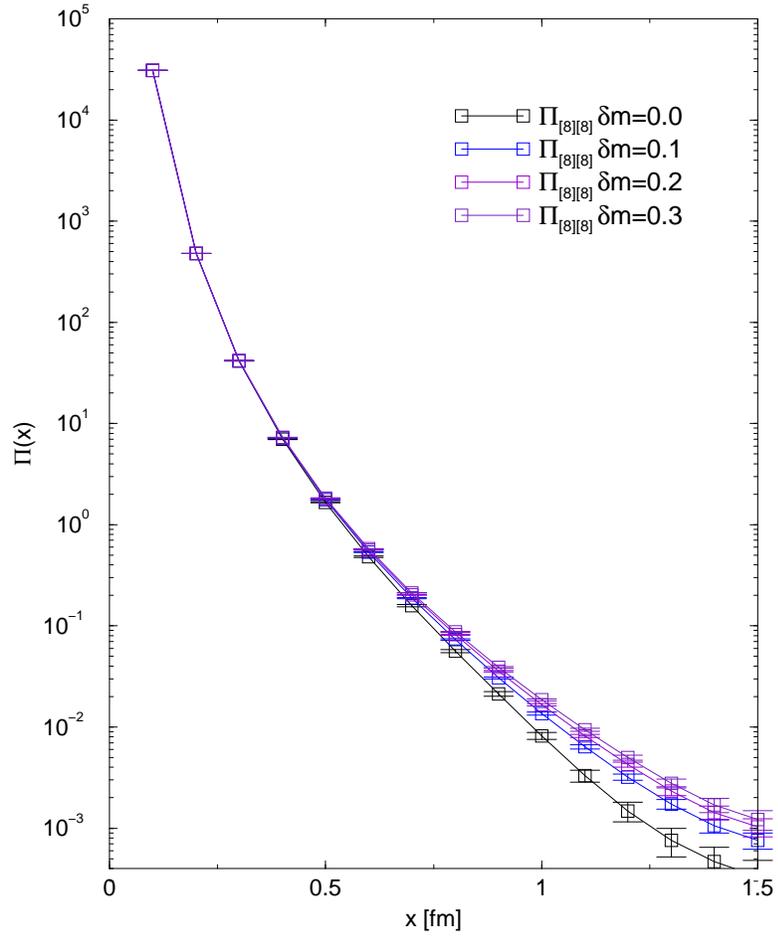}
\end{center}
\caption{Correlation functions in the color-octet/flavor-octet 
([8][8]) channel for different values of the explicit flavor
symmetry breaking parameter $\delta m=0.0,0.1,0.2,0.3\,{\rm 
fm}^{-1}$. The correlators were calculated in a quenched 
instanton ensemble. }
\end{figure}


\begin{thebibliography}{20}

\bibitem{Wetterich:1999vd}
C.~Wetterich,
Phys.\ Lett.\ B {\bf 462}, 164 (1999)
[hep-th/9906062].

\bibitem{Wetterich:2000pp}
C.~Wetterich,
hep-ph/0008150.

\bibitem{Wetterich:2000ky}
C.~Wetterich,
hep-ph/0011076.

\bibitem{Berges:2000dh}
J.~Berges and C.~Wetterich,
hep-ph/0012311.

\bibitem{Srednicki:1981cu}
M.~Srednicki and L.~Susskind,
Nucl.\ Phys.\ B {\bf 187}, 93 (1981).

\bibitem{Iijima:1981tt}
B.~Iijima and R.~L.~Jaffe,
Phys.\ Rev.\ D {\bf 24}, 177 (1981).

\bibitem{Alford:1999mk}
M.~Alford, K.~Rajagopal and F.~Wilczek,
Nucl.\ Phys.\ B {\bf 537}, 443 (1999)
[hep-ph/9804403].

\bibitem{Schafer:1999ef}
T.~Sch{\"a}fer and F.~Wilczek,
Phys.\ Rev.\ Lett.\ {\bf 82}, 3956 (1999)
[hep-ph/9811473].

\bibitem{Bailin:1984bm}
D.~Bailin and A.~Love,
Phys.\ Rept.\ {\bf 107}, 325 (1984).

\bibitem{Rapp:1998zu}
R.~Rapp, T.~Sch{\"a}fer, E.~V.~Shuryak and M.~Velkovsky,
Phys.\ Rev.\ Lett.\ {\bf 81}, 53 (1998)
[hep-ph/9711396].

\bibitem{Alford:1998zt}
M.~Alford, K.~Rajagopal and F.~Wilczek,
Phys.\ Lett.\ B {\bf 422}, 247 (1998)
[hep-ph/9711395].

\bibitem{Rapp:2000qa}
R.~Rapp, T.~Sch{\"a}fer, E.~V.~Shuryak and M.~Velkovsky,
Annals Phys.\ {\bf 280}, 35 (2000)
[hep-ph/9904353].

\bibitem{Schafer:2000fe}
T.~Sch{\"a}fer,
Nucl.\ Phys.\ B {\bf 575}, 269 (2000)
[hep-ph/9909574].

\bibitem{Pisarski:2000gq}
R.~D.~Pisarski,
Phys.\ Rev.\ C {\bf 62}, 035202 (2000)
[nucl-th/9912070].

\bibitem{Schafer:1998wv}
T.~Sch{\"a}fer and E.~V.~Shuryak,
Rev.\ Mod.\ Phys.\ {\bf 70}, 323 (1998)
[hep-ph/9610451].

\bibitem{Diakonov:1995ea}
D.~Diakonov,
hep-ph/9602375.

\bibitem{Shifman:1979bx}
M.~A.~Shifman, A.~I.~Vainshtein and V.~I.~Zakharov,
Nucl.\ Phys.\ B {\bf 147}, 385 (1979).

\end{thebibliography}
\end{document}